\documentclass[12pt]{article} 
 \begin{document}
\baselineskip=18pt
\title{Hydrodynamic approach to time-dependent density functional
theory; response properties of metal clusters}
\author{Arup Banerjee and Manoj K. Harbola\\
Laser Programme, Centre for Advanced Technology,\\
Indore 452013, INDIA.}
\maketitle

\begin{abstract}
Performing electronic structure calculations for large systems, such as 
nanoparticles or metal clusters, via orbital
based Hartree-Fock or Kohn-Sham theories is computationally demanding. To
study such systems, therefore, we have taken recourse to the hydrodynamic
approach to time-dependent density-functional theory. In this paper we develop
variation-perturbation method within this theory in terms of the particle
and current densities of a system. We then apply this to study the linear and
nonlinear response properties of alkali metal clusters within the spherical
jellium background model.
\end{abstract}

\newpage
\section*{I Introduction}
The hydrodynamic analogy of quantum mechanics was first explored by Madelung
\cite{madelung} who transformed the single particle Schrodinger equation 
into a pair of hydrodynamical equations. The theory views \cite{lund,deb1} the
electron cloud as a classical fluid moving under the action of classical 
Coulomb forces augmented by the forces of quantum origin. The basic 
dynamical variables of this theory are the particle and current densities 
which satisfy two fluid dynamical equations, namely, the continuity and an 
Euler type equation. The work of Madelung was followed by Bloch's attempt 
\cite{bloch} to develop hydrodynamical theory for many-electron systems
within the realm of Thomas-Fermi (TF) theory \cite{thomas,fermi,march}. 
Although then proposed without any rigorous foundation, the theory can now 
be derived \cite{harbola} 
from the equations of time-dependent density-functional theory (TDDFT)
\cite{gross,casida}. It is based on the assumption
that the dynamics of many-electron system can be described by considering 
it as a fluid of density $\rho({\bf r},t)$ and a velocity field 
${\bf v}({\bf r},t)$ which is assumed to be curl free 
(that is ${\bf v}({\bf r},t) = -{\bf\nabla}
S({\bf r},t)$, where $S({\bf r},t)$ is scalar velocity potential).

Using $\rho({\bf r},t)$ and $S({\bf r},t)$ as conjugate variables Bloch derived
two fluid dynamical equations: the continuity equation
\begin{equation}
\frac{\partial\rho}{\partial t} + {\bf\nabla}\cdot(\rho{\bf\nabla}S) = 0,
\label{1}
\end{equation}
and the Euler equation
\begin{equation}
\frac{\partial S}{\partial t} = \frac{1}{2}|{\bf\nabla}S|^{2} + 
\frac{\delta T_{0}}{\delta\rho} +  v_{ext}({\bf r},t) + 
\int\frac{\rho({\bf r'},t)}{|{\bf r} - {\bf r'}|}d{\bf r'}.
\label{2}
\end{equation}
Here $T_{0}$ is the TF kinetic energy (KE) functional and $v_{ext}({\bf r},t)$
represents the external potential. These equations were subsequently used to
study photoabsorption cross section and collective excitation of 
atoms\cite {ball}, collective excitations \cite{walecka} and plasmons of 
metal clusters \cite{rupin} and surface plasmons \cite{bennet,egui} in metals.

Bloch's theory also formed the basis of initial attempts by Ying \cite{ying}
to extend density-functional theory (DFT) to include time-dependent (TD) 
external potentials. He did this by replacing the TF KE term $T_{0}$ by a 
general functional $G[\rho]$ consisting of the KE and the exchange-correlation
(XC) contribution to the total energy. In Ying's work it is implicit that, 
like in the static DFT, a universal functional $G[\rho({\bf r},t)]$ can be 
written for the TD problem. The ad-hoc nature of Bloch's theory and its 
extension
by Ying was removed by the pioneering works of Deb and Ghosh \cite{deb2}, 
Bartolotti \cite{bartolotti} and Runge and Gross \cite{runge}. 
Runge and Gross rigorously proved the existence of a
Hohenberg-Kohn \cite{hk} like theorem for TD potentials, and showed that
the TD density $\rho({\bf r},t)$ can be 
determined by solving the hydrodynamical equations
\begin{equation}
\frac{\partial\rho}{\partial t} + {\bf\nabla}\cdot{\bf j} = 0,
\label{3}
\end{equation}
which is the continuity equation, and the Euler's equation
\begin{equation}
\frac{\partial{\bf j}}{\partial t} = {\bf P}[\rho({\bf r},t)].
\label{4}
\end{equation}
Here ${\bf P}$ is the three-component density-functional of Runge and Gross
and the vector ${\bf j}$ is the current density corresponding to the many-body
wavefunction $\Psi({\bf r}_{1},\cdots,{\bf r}_{N};t)$. An explicit expression 
for ${\bf P}[\rho({\bf r},t)]$ in terms of the wavefunction has recently been 
given \cite{harbola} using the TD differential virial theorem\cite{sahni}.

Although there have been some calculations, as mentioned above, in the
past by employing
hydrodynamic theory, its full potential remains unexplored. This is 
evidently because with the increasing computing resources one can perform \cite{yanno}
orbital based calculations like 
TD Hartee-Fock (HF) or TD Kohn-Sham (KS) with relative ease. Recently,
however, hydrodynamical theory is being applied in situations where such 
orbital based calculations are still computationally difficult to implement. 
One such example is systems
which contain thousands of atom such as nanoparticles and clusters. 
In these systems hydrodynamical theory becomes the method of choice. Thus the
theory has been applied to study photoabsorption cross-section of 
metal particles \cite{domps}, collective \cite{zaremba1} and
magnetoplasmon excitations \cite{zaremba2} of confined electronic systems and also
to study the interaction of strong laser light with atomic systems 
\cite{deb3,dey,gross1}. Besides the computational ease offered by it, one
is also tempted to work within the hydrodynamic formulation because it
provides an intuitively appealing approach to the time dependent many-body
problem.  

In this paper we develop perturbation theory within the hydrodynamic formalism
to calculate linear and nonlinear response properties of large systems. 
The motivation for this comes from our experience with the calculation of 
static response properties \cite{baner1,baner2} employing density based 
perturbation theory \cite{harbola2} within the  Hohenberg-Kohn formalism of 
DFT. The density based method reduces the 
numerical effort required for such calculations substantially while leading to
reasonably accurate results \cite{baner1,baner2} for the response properties. In the same manner, 
the hydrodynamical approach proves to be useful for calculating frequency
dependent response properties of extended systems for which orbital based
theories become quite difficult to implement because of the large number of 
orbitals involved.

The work presented in this paper is divided into two parts: First, we derive
the generalized Bloch type equation using the concept of time-averaged
energy (quasi-energy) of an electronic system subjected to a TD periodic
field. For calculating optical response properties we then develop 
variation-perturbation (VP) theory in terms of the quasi-energy using particle
and current densities as the basic variables. This is presented in section II.
The perturbation theory developed here proceeds along the lines of density
based stationary-state perturbation theory \cite{harbola2} by making use of 
the stationary nature of the time-averaged energy with respect to 
$\rho({\bf r},t)$ and $S({\bf r},t)$. In the second part
we demonstrate the applicability of the perturbation theory developed here
by calculating frequency dependent linear and nonlinear polarizabilities of
inert gas atoms (section III) and comparing our numbers with the
standard results obtained from the wavefunctional approach. Having 
demonstrated the accuracy of hydrodynamical approach, we then apply it to 
calculate frequency dependent response properties of metal clusters with 
number of atoms up to 5000 using the spherical jellium background model 
(SJBM).
\section*{II Variation perturbation method in hydrodynamical theory}
\subsection*{A. Time-averaged energy}
The central quantity around which the VP theory is developed in the static
case is the ground-state energy of the system. For periodic TD hamiltonians,
this role is played by the time-averaged energy or quasi-energy \cite{sambe}.
Therefore, in the following we first derive an expression for the 
quasi-energy as a functional of particle and current densities and show that 
it obeys stationarity for the correct solutions of these functions. As 
expected from the work of Runge and Gross \cite{runge},  stationarity with respect to the 
density leads to the equation of motion for the density. In addition, 
variation with respect to the current density gives the continuity equation.

We begin with the TD Schrodinger equation for the many-body wave function
$\Psi({\bf r}_{1},\cdots,{\bf r}_{N};t)$ given by
\begin{equation}
\left ( H(t) - \imath\frac{\partial}{\partial t}\right )\Psi ({\bf r},t) = 0
\label{5}
\end{equation}
where
\begin{equation}
H(t) =  \hat{T} + \hat{V}_{ee} + \hat{V}(t).
\label{6}
\end{equation}
In the above equation $\hat{T}$ and $\hat{V}_{ee}$ are the kinetic and 
electron-electron interaction energy operators, respectively, and 
$\hat{V}(t)$ denotes the TD external potential containing both the nuclear 
and the applied potential. For periodic hamiltonians, that is
\begin{equation}
H(t + T) = H(t),
\label{7}
\end{equation}
where T is the time period, in accordance with Floquet's theorem there 
exists a solution $\Psi({\bf r},t)$
of the form
\begin{equation}
\Psi({\bf r},t) = \Phi({\bf r},t)e^{-\imath Et}
\label{8}
\end{equation}
where $\Phi({\bf r}_{1},\cdots,{\bf r}_{N};t)$ is also periodic in time with 
time period T, i.e., $\Phi(t+T) = \Phi(t)$. Such a state has been termed as 
the steady-state of the system  with $E$ being the corresponding quasi-energy.
The equation of motion for $\Phi$ is easily seen to be
\begin{equation}
\left ( H(t) - \imath\frac{\partial}{\partial t}\right )\Phi ({\bf r},t) = 
E\Phi ({\bf r},t). 
\label{9}
\end{equation}
The corresponding expression for the quasi-energy is the time averaged expectation
value
\begin{equation}
E[\Phi] = \left\{\langle\Phi|H(t) - \imath\frac{\partial}{\partial t}|\Phi\rangle\right\}.
\label{10}
\end{equation}
The curly bracket in Eq.(\ref{10}) denotes the time averaging over one period T defined as
\begin{equation}
\left\{fg\right\} = \frac{1}{T}\int_{0}^{T}f^{*}(t)g(t)dt
\label{11}
\end{equation}
The quasi-energy represents the average energy of induction \cite{langhoff} 
of a system subjected to a TD
potential as is easily seen by the TD Hellmann-Feynman theorem \cite{parr}.

To convert Eq.(\ref{10}) into its hydrodynamical counterpart we decompose the complex
steady-state many-body wavefunction in polar form, so that
\begin{equation}
\Phi({\bf r},t) = \chi({\bf r},t)e^{\imath S({\bf r},t)}
\label{12}
\end{equation}
where both $\chi$ and $S$ are real functions of ${\bf r}_{1}, {\bf r}_{2},\cdots,{\bf r}_{N}$
and are periodic in time with time period T. Further, $S({\bf r},t)$ also
has a purely TD component which integrates to zero over time period T (for
detail see Ref.\cite{langhoff}). 
Note that, $S$ is zero for the ground-state of the system.
By substituting Eq.(\ref{12}) in Eq.(\ref{10}), the
expression for the average energy becomes
\begin{equation}
E[\chi ,S] = \left\{\langle\chi|\hat{T}' + \hat{V}_{ee} + \frac{\partial S}{\partial t}|\chi
\rangle - \imath\langle\chi|\frac{\partial\chi}{\partial t}\rangle\right\}
\label{13}
\end{equation}
where
\begin{equation}
\hat{T}' = \sum_{i}\left(-\frac{1}{2}\nabla_{i}^{2} + \frac{1}{2}|{\bf\nabla}_{i}S|^{2}\right)
\label{14}
\end{equation}
Since the periodicity and reality of $\chi({\bf r},t)$ implies that
\begin{eqnarray}
\left\{\langle\chi|\frac{\partial\chi}{\partial t}\rangle\right\} &  = & \frac{1}{2T}\int_{0}^{T}dt\int
\frac{\partial}{\partial t}(\chi^{*}\chi)d{\bf r}\nonumber \\
& = & 0,
\label{15}
\end{eqnarray}
the quasi-energy is given as
\begin{equation}
E[\chi,S] = \left\{\langle\chi|-\sum_{i}\frac{1}{2}\nabla_{i}^{2} + \hat{V}_{ee} + 
\hat{V}_{ext} + \sum_{i}\frac{1}{2}
({\bf\nabla}_{i}S)^{2} + \frac{\partial S}{\partial t}|\chi\rangle\right\}
\label{16}
\end{equation}
Now by invoking the Runge-Gross theorem \cite{runge}, it can be written as a
functional of the density alone. However, in the hydrodynamical formulation
the density and the velocity potential $S$ are treated as independent variables which means 
that the energy above is a functional of these two quantities. An 
advantage of this decoupling of the density and $S$ is that one does 
not have to know the functional 
dependence of the $S$ on the density. Further, this facilitates 
approximating the expectation value $\langle\chi|-\frac{1}{2}
\nabla^{2}|\chi\rangle$ as a functional of the density by the KE functionals
well studied in static DFT. Evidently, the first three terms
of the equation above can be represented by a functional of TD density as
\begin{equation}
\left\{F[\rho({\bf r},t)] + \int\left [v_{0}({\bf r}) + v_{app}({\bf r},t)
\right]\rho({\bf r},
t)d{\bf r}\right\},
\label{17}
\end{equation}
where $v_{0}({\bf r})$ represents the static external potential and 
TD part of the potential is represented by $v_{app}({\bf r},t)$. 
This is because changes in $S$ do not affect their values.
The universal functional $\left\{F[\rho({\bf r},t)]\right\}$ given by
\begin{equation}
\left\{F[\rho({\bf r},t)]\right\} = \left\{T_{s}[\rho({\bf r},t)]\right\}
 + \left\{E_{H}[\rho({\bf r},t)]\right\} + \left\{E_{xc}[\rho({\bf r},t)]
\right\}.  
\label{18}
\end{equation}
Here $\left\{T_{s}[\rho({\bf r},t)]\right\}$, $\left\{E_{H}[\rho({\bf r},t)]
\right\}$ and $\left\{E_{xc}[\rho({\bf r},t)]\right\}$ represent the 
time-averaged KE, Hartree energy and exchange and correlation (XC) energy 
functionals respectively for the TD system. The TD particle density is given by
\begin{equation}
\rho({\bf r},t) = \int\chi^{*}({\bf r}, {\bf r}_{2},\cdots,{\bf r}_{N},t)
\chi({\bf r},{\bf r}_{2},\cdots,{\bf r}_{N},t)d{\bf r}_{2}\cdots d{\bf r}_{N}
\label{19}
\end{equation}
So far we have written the first three terms in terms of the particle density,
a quantity defined in 3D configuration space. The last two terms representing
the current still have all the co-ordinates of the configuration space in 
them. As such any equation involving $S({\bf r}_{1},\cdots,{\bf r}_{N};t)$ 
can not be projected on to 3D space. To do this one needs to consider some 
approximate form for the phase $S$. One such approximation for $S$ which is
generally employed \cite{deb1} is that it can be written as the sum of single 
particle phases, that is
\begin{equation}
S({\bf r}_{1},\cdots,{\bf r}_{N};t) = \sum_{i=1}^{N}S({\bf r}_{i};t)
\label{20}
\end{equation}
with the same function $S$ representing each electron. This approximation is 
equivalent to 
assuming the velocity field of the electron fluid to be curl free as was
done by Bloch \cite{bloch} in deriving Eq.(\ref{2}). With this approximation 
the average energy functional of Eq.(\ref{16}) is given as
\begin{eqnarray}
E[\rho,S] & = & \left\{F[\rho({\bf r},t)] + \int\left(v_{0}({\bf r}) + v_{app}
({\bf r},t)\right)
\rho({\bf r},t)d{\bf r}\right. \nonumber \\ 
& + & \left. \frac{1}{2}\int\rho({\bf r},t)({\bf\nabla}S)^{2}d{\bf r} + 
\int\frac{\partial S}{\partial t}\rho({\bf r},t)d{\bf r}\right\}
\label{21}
\end{eqnarray}
This is the expression for the quasi-energy of a many electron system (under 
the approximation made above) interacting with a TD periodic potential. Since 
the purely TD component of $S({\bf r},t)$  is the same as the phase of the 
wavefunction, it does not contribute to the energy.  In Eq.\ref{21} this is
ensured by $\rho({\bf r},t)$ integrating to a fixed number of electrons at
all times. We therefore drop it and work with only the co-ordinate dependent 
component of $S({\bf r},t)$. This is similar to
separating out the overall phase of TD wavefunction \cite{langhoff,sambe}
in the TD perturbation theory.

We now demonstrate the variational nature of $E[\rho,S]$ with respect to 
$\rho$ and $S$. Making $E[\rho,s]$ stationary with respect to $\rho$ and 
$S$ gives the Euler equation
\begin{equation}
\mu(t) = -\frac{\partial S}{\partial t} + \frac{1}{2}({\bf\nabla}S)^{2} + 
v_{0}({\bf r}) + v_{app}({\bf r},t) + \frac{\delta F}{\delta\rho}
\label{22}
\end{equation}
where $\mu(t)$ is the Lagrange-multiplier ensuring that $\rho({\bf r},t)$ 
integrates to the correct number of electrons at each instant of time, and 
the continuity  equation
\begin{equation}
\frac{\partial\rho}{\partial t} + {\bf\nabla}\cdot(\rho{\bf\nabla}S) = 0,
\label{23}
\end{equation}
respectively. Eq.(\ref{22}) is the same as that proposed by Ying\cite{ying}.
As such if 
$F[\rho]$ is approximated by the TF functional, it gives the Bloch's 
hydrodynamical equation correctly. Further, for time independent
hamiltonians it correctly reduces to the Euler equation of static DFT.
All these facts demonstrate the variational nature of $E[\rho,S]$ with 
respect to $\rho$ and $S$.  Employing this we now develop the VP method in 
terms of the particle and the current densities. We show that the $(2n + 1)$ 
theorem and its variational corollary is satisfied in terms of
these variables.
\subsection*{B. Perturbation theory}
To develop perturbation theory we assume that $v_{ext}({\bf r},t)$ is 
relatively weak and under its action the ground-state density 
$\rho^{(0)}({\bf r})$ changes to $\rho^{(0)}({\bf r}) + 
\Delta\rho({\bf r},t)$ and the velocity potential changes to
$S^{(0)}({\bf r}) + \Delta S({\bf r},t)$. The particle density change 
$\Delta\rho$ satisfies the normalization condition
\begin{equation}
\int\Delta\rho({\bf r},t)d{\bf r} = 0.
\label{24}
\end{equation}
However no such condition is required for the change in the velocity 
potential $\Delta S$. The changes $\Delta\rho$ and $\Delta S$ are expanded in 
perturbation series as
\begin{eqnarray}
\Delta\rho & = & \sum_{j}\rho^{(j)} \nonumber \\
\Delta S & = & \sum_{j}S^{(j)}, 
\label{25}
\end{eqnarray}
where $\rho^{(j)}$ and $S^{(j)}$ correspond to the $jth$ order terms in the
perturbation parameter. The energy corresponding to $\rho^{(0)} + 
\Delta\rho$ and $S^{(0)} + \Delta S$ is given by
\vspace{0.1in}
\begin{eqnarray}
E[\rho^{(0)} + \Delta\rho,S^{(0)} + \Delta S] & = & \left\{F[\rho^{(0)} + 
\Delta\rho] + \int(v_{0} + 
v_{app})(\rho^{(0)} + \Delta\rho)d{\bf r} \right. \nonumber \\
& + & \left. \frac{1}{2}\int{\bf\nabla}(S^{(0)} + \Delta S)
\cdot{\bf\nabla}(S^{(0)} + \Delta S)
(\rho^{(0)} + \Delta\rho)d{\bf r} \right. \nonumber \\
& + & \left. \int\frac{\partial (S^{(0)} + \Delta S)}{\partial t}
(\rho^{(0)} + \Delta\rho)d{\bf r} \right\}
\label{26}
\end{eqnarray}
Using Eq.({\ref{26}) we now obtain the energy changes to different orders in 
perturbation parameter employing an approach identical to the one adopted in 
Ref.\cite{harbola2} for time independent
density based perturbation theory. The resulting expressions for average 
energies to different orders are:
\begin{equation}
E^{(1)} = \left\{\int v_{app}^{(1)}({\bf r},t)\rho^{(0)}({\bf r})d{\bf r}
\right\},
\label{27}
\end{equation}
\begin{eqnarray}
E^{(2)}& =& \left\{\frac{1}{2}\int\frac{\delta^{2} F[\rho]}{\delta\rho
({\bf r},t)\delta\rho({\bf r'},t)}
\rho^{(1)}({\bf r},t)\rho^{(1)}({\bf r'},t)d{\bf r}d{\bf r'}   
+ \int v_{app}^{(1)}({\bf r},t)\rho^{(1)}({\bf r})d{\bf r}\right. \nonumber \\
& + & \left.\int\frac{\partial S^{(1)}({\bf r},t)}{\partial t}\rho^{(1)}
({\bf r},t)d{\bf r} + \frac{1}{2}\int ({\bf\nabla}S^{(1)}\cdot{\bf\nabla}
S^{(1)})\rho^{(0)}({\bf r})d{\bf r}\right\},
\label{28}
\end{eqnarray}
\begin{eqnarray}
E^{(3)}& = &\left\{\frac{1}{6}\int\frac{\delta^{3} F[\rho]}
{\delta\rho({\bf r},t)\delta\rho({\bf r'},t)
\delta\rho({\bf r''},t)}\rho^{(1)}({\bf r},t)\rho^{(1)}
({\bf r'},t)\rho^{(1)}({\bf r''},t)
d{\bf r}d{\bf r'}d{\bf r''}\right. \nonumber \\
& + & \left.\int({\bf\nabla}S^{(1)}\cdot{\bf\nabla}S^{(1)})\rho^{(1)}
({\bf r},t)d{\bf r}\right\},
\label{29}
\end{eqnarray}
\begin{eqnarray}
E^{(4)}& = &\left\{\frac{1}{2}\int\frac{\delta^{2} 
F[\rho]}{\delta\rho({\bf r},t)\delta\rho({\bf r'},t)}
\rho^{(2)}({\bf r},t)\rho^{(2)}({\bf r'},t)d{\bf r}d{\bf r'} \right. 
\nonumber \\
& + & \left.\frac{1}{6}\int\frac{\delta^{3} F[\rho]}{\delta\rho({\bf r},t)
\delta\rho({\bf r'},t)\delta\rho({\bf r''},t)}\rho^{(2)}({\bf r},t)
\rho^{(1)}({\bf r'},t)\rho^{(1)}({\bf r''},t)
d{\bf r}d{\bf r'}d{\bf r''}\right. \nonumber \\
& + & \left.\frac{1}{24}\int\frac{\delta^{4} F[\rho]}{\delta\rho({\bf r},t)
\delta\rho({\bf r'},t)\delta\rho({\bf r''},t)\delta\rho({\bf r'''},t)}
\rho^{(1)}({\bf r},t)\rho^{(1)}({\bf r'},t)
\rho^{(1)}({\bf r''},t)\rho^{(1)}({\bf r'''},t)\right.\nonumber \\
&\times& \left. d{\bf r}d{\bf r'}d{\bf r''}d{\bf r'''}\right. \nonumber \\
& + & \left.\frac{1}{2}\int ({\bf\nabla}S^{(2)}\cdot{\bf\nabla}S^{(2)})
\rho^{(0)}({\bf r})d{\bf r} +
\frac{1}{2}\int ({\bf\nabla}S^{(1)}\cdot{\bf\nabla}S^{(1)})
\rho^{(2)}({\bf r},t)d{\bf r}\right.
\nonumber \\
&+& \left.\int ({\bf\nabla}S^{(1)}\cdot{\bf\nabla}S^{(2)})\rho^{(1)}
({\bf r},t)d{\bf r} + \int\frac{\partial S^{(2)}}{\partial t}\rho^{(2)}
({\bf r},t)d{\bf r} \right\}
\label{30}
\end{eqnarray} 
and
\begin{eqnarray}
E^{(5)}& = &\left\{\frac{1}{2}\int\frac{\delta^{3} F[\rho]}
{\delta\rho({\bf r},t)\delta\rho({\bf r'},t)
\delta\rho({\bf r''},t)}\rho^{(2)}({\bf r},t)\rho^{(2)}
({\bf r'},t)\rho^{(1)}({\bf r''},t)
d{\bf r}d{\bf r'}d{\bf r''}\right. \nonumber \\
& + & \left.\frac{1}{24}\int\frac{\delta^{4} F[\rho]}{\delta
\rho({\bf r},t)\delta\rho({\bf r'},t)
\delta\rho({\bf r''},t)\delta\rho({\bf r'''},t)}\rho^{(1)}
({\bf r},t)\rho^{(1)}({\bf r'},t)
\rho^{(1)}({\bf r''},t)\rho^{(2)}({\bf r'''},t)\right.\nonumber \\
& &\times \left. d{\bf r}d{\bf r'}d{\bf r''}d{\bf r'''}\right. \nonumber \\
& + & \left.\frac{1}{120}\int\frac{\delta^{5} F[\rho]}
{\delta\rho({\bf r},t)\delta\rho({\bf r'},t)
\delta\rho({\bf r''},t))\delta\rho({\bf r'''},t)\delta\rho({\bf r''''},t)}
\rho^{(1)}({\bf r},t)\rho^{(1)}({\bf r'},t)\rho^{(1)}
({\bf r''},t)\right.\nonumber \\
& &\times\left.\rho^{(1)}({\bf r'''},t)\rho^{(1)}({\bf r''''},t)
d{\bf r}d{\bf r'}d{\bf r''}d{\bf r'''}d{\bf r''''}\right. \nonumber \\
&+& \left.\int ({\bf\nabla}S^{(2)}\cdot{\bf\nabla}S^{(2)})
\rho^{(1)}({\bf r},t)d{\bf r} + \int ({\bf\nabla}S^{(1)}
\cdot{\bf\nabla}S^{(2)})\rho^{(2)}({\bf r},t)d{\bf r}\right\} 
\label{31}
\end{eqnarray} 
In deriving these equations (Eq.(\ref{27})-(\ref{31})) we have made 
use of the fact that for the ground-state $S^{(0)}=0$ which implies
that the current density $\rho^{(0)}{\bf\nabla}S^{(0)}$ and the
time-derivative $\frac{\partial S^{(0)}}{\partial t}$ vanish for the 
ground-state. In addition we also use the first-order
\begin{equation}
\frac{\partial\rho^{(1)}}{\partial t} + {\bf\nabla}\cdot(\rho^{(0)}{\bf\nabla}
S^{(1)}) = 0
\label{34}
\end{equation}
\begin{equation}
\mu^{(1)}(t) = -\frac{\partial S^{(1)}}{\partial t} + v_{app}^{(1)}({\bf r},t)
+ \frac{1}{2}\int\frac{\delta^{2}F[\rho]}{\delta\rho({\bf r},t)\delta
\rho({\bf r'},t)}\rho^{(1)}({\bf r'},t)d{\bf r'}
\label{35}
\end{equation}
and the second-order
\begin{equation}
\frac{\partial\rho^{(2)}}{\partial t} + {\bf\nabla}\cdot(\rho^{(0)}
{\bf\nabla}S^{(2)}) + {\bf\nabla}\cdot(\rho^{(1)}{\bf\nabla}S^{(1)}) = 0
\label{36}
\end{equation}
\begin{eqnarray}
\mu^{(2)}(t) & = & -\frac{\partial S^{(2)}}{\partial t} + \frac{1}{2}({\bf\nabla}S^{(1)}\cdot
{\bf\nabla}S^{(1)})\nonumber \\
& + &\frac{1}{2}\int\frac{\delta^{2}F[\rho]}{\delta\rho({\bf r},t)\delta\rho({\bf r'},t)}
\rho^{(2)}({\bf r'},t)d{\bf r'}\nonumber \\
& + &\frac{1}{2}\int\frac{\delta^{3}F[\rho]}{\delta\rho({\bf r},t)\delta\rho({\bf r'},t)
\delta\rho){\bf r''},t)}\rho^{(1)}({\bf r'},t)\rho^{(1)}({\bf r''},t)
d{\bf r'}d{\bf r''}
\label{37}
\end{eqnarray}
continuity and Euler equations obtained by expanding Eqs.(\ref{21}) and 
(\ref{22}). Expressions for the average energies (Eqs.(\ref{27})-(\ref{31})) 
clearly demonstrates the $(2n + 1)$ rule of perturbation theory. Energies 
up to order 3 are determined completely by $\rho^{(1)}$ and 
$S^{(1)}$. Similarly $\rho^{(2)}$ and $S^{(2)}$ give energy up to the
fifth-order. This is the $(2n + 1)$ theorem of hydrodynamic perturbation 
theory in terms of the particle and the current densities. Moreover, 
even-order corollary of this theorem also holds true. Thus making
E$^{(2)}$ stationary with respect to $S^{(1)}$ and $\rho^{(1)}$, leads to 
Eqs.(\ref{34}) and (\ref{35}). Similarly stationarity of E$^{(4)}$ with 
respect to $\rho^{(2)}$ and $S^{(2)}$ gives the correct perturbation 
equations (Eqs.(\ref{36}) and (\ref{37})) for $\rho^{(2)}$ and $S^{(2)}$. The 
stationary nature of the even-order energies
gives a variational method to obtain approximate solutions for the 
corresponding induced densities and currents.

With this we complete the development of VP method in terms of particle and 
current densities in hydrodynamic formulation of TDDFT. In the next section 
we demonstrate the applicability of this theory by calculating the frequency 
dependent linear and nonlinear polarizabilities
of inert gas atoms and comparing the results obtained with their 
wavefunctional counterparts. We then apply the formalism to calculate 
frequency dependent polarizabilities and plasmon
frequencies of alkali metal clusters of large sizes.
\section*{III. Application of hydrodynamical formalism}
To demonstrate the applicability of the formalism developed above we begin 
this section with the calculation of frequency dependent linear and nonlinear
polarizabilities of inert gas atoms. In the present formulation we can 
calculate the nonlinear polarizabilities corresponding to the degenerate 
four wave mixing (DFWM) and DC-Kerr \cite{shen} effect which are directly
related to the fourth-order energy changes. On the other hand, unlike the 
orbital based Kohn-Sham approach, $(2n + 1)$ theorem cannot be exploited to 
calculate \cite{baner4,harbola3,baner3} the coefficients for the third-harmonic 
generation and electric field induced second harmonic generation processes 
from only the second-order induced densities. In the following we will present
the results for the nonlinear coefficients corresponding to the DFWM 
process only.
 
As pointed out earlier, we perform our calculations using the variational 
property of the even-order energies. To this end we choose an appropriate 
variational form for the induced particle and current densities. For an 
applied potential of the form
\begin{equation}
v_{app}^{(1)}({\bf r},t) = v_{app}^{(1)}({\bf r})\cos\omega t
\label{38}
\end{equation}
with the spatial part given by
\begin{equation}
v_{app}^{(1)}({\bf r}) = {\cal E}r\cos\theta
\label{39}
\end{equation}
where ${\cal E}$ is amplitude of the applied field, the time dependence of $\rho$ and $S$ 
at various orders can easily be inferred from  Eqs(\ref{34})-(\ref{36}) as
\begin{eqnarray}
\rho^{(1)}({\bf r},t)& = & \rho^{(1)}({\bf r})\cos\omega t\nonumber \\
\rho^{(2)}({\bf r},t)& = & \rho_{2}^{(2)}({\bf r})\cos2\omega t + \rho_{0}^{(2)}({\bf r})
\label{40}
\end{eqnarray}
and
\begin{eqnarray}
S^{(1)}({\bf r},t) & = & S^{(1)}({\bf r})\sin\omega t\nonumber \\
S^{(2)}({\bf r},t) & = & S^{(2)}({\bf r})\sin 2\omega t.
\label{41}
\end{eqnarray}
Note that unlike the second-order particle density, the corresponding current
has no constant term. This is consistent with the fact that the current 
arises due to the flow of electrons which causes the density to be time 
dependent. The spatial part of the induced particle and current densities are
determined variationally by minimizing the appropriate even-order energies. 
For this purpose we choose the forms of the induced particle densities 
similar to the ones used previously \cite{baner1,baner2} for the calculation 
of static response properties. These are
\begin{equation}
\rho^{(1)}({\bf r}) = \Delta_{1}(r)\rho^{(0)}({\bf r})\cos\theta
\label{42}
\end{equation}
and
\begin{eqnarray}
\rho_{2}^{(2)}({\bf r})& = & \left (\Delta_{2}^{2}(r) + \Delta_{3}^{2}(r)\cos^{2}\theta\right )
\rho^{(0)}({\bf r}) + \lambda_{2}\rho^{(0)}({\bf r}) \nonumber \\
\rho_{0}^{(2)}({\bf r})& = & \left (\Delta_{2}^{0}(r) + \Delta_{3}^{0}(r)\cos^{2}\theta\right )
\rho^{(0)}({\bf r}) + \lambda_{0}\rho^{(0)}({\bf r}) 
\label{43}
\end{eqnarray}
with
\begin{equation}
\Delta_{i}(r) = \sum_{j}a_{j}^{i}r^{j}
\label{44}
\end{equation}
where $\rho^{(0)}({\bf r})$ is the ground-state density, $a_{j}^{i}$ are the 
variational parameters and $\lambda$s are fixed by the normalization
condition for $\rho^{(2)}$. To ensure satisfaction of the normalization 
condition at all times, $\rho_{2}^{(2)}$ and $\rho_{0}^{(2)}$ are each 
normalized separately. These forms for the induced particle densities are 
motivated by the exact solutions \cite{coulson,swell} for the hydrogen atom 
in a static field and have been shown \cite{baner1,baner2} to lead to 
accurate static polarizabilities. On the basis of the 
continuity equation at each order and Eqs.(\ref{42})-(\ref{44}), we choose
\begin{equation}
S^{(1)}({\bf r}) = \Delta_{1}^{s}(r)\cos\theta
\label{45}
\end{equation}
and
\begin{equation}
S^{(2)}({\bf r}) = \left (\Delta_{2}^{s}(r) + \Delta_{3}^{s}(r)
\cos^{2}\theta\right ),
\label{46}
\end{equation}
where
\begin{equation}
\Delta_{i}^{s}(r) = \sum_{j}b_{j}^{i}r^{j}
\label{47}
\end{equation} 
with $b_{j}^{i}$ being the variational parameters to be determined by 
minimizing the average energy of respective orders.   

Application of hydrodynamical equations also requires approximating the 
KE and XC energy functionals. Based on our experience with the calculation of 
static linear and nonlinear polarizabilities we approximate them by their static forms.
Thus for the KE, we use the von Weizsacker 
\cite{von} functional.
\begin{equation}
T_{s}[\rho] = \frac{1}{8}\int\frac{{\bf\nabla}\rho\cdot{\bf\nabla}\rho}
{\rho}d{\bf r}
\label{48}
\end{equation}
For a discussion on the rationale behind choosing this
functional, we refer the reader to the literature 
\cite{baner1,baner2,jones1,jones2}. For the exchange energy functional we use
the adiabatic local-density approximation (ALDA) given by the Dirac exchange 
functional \cite{dirac}
\begin{eqnarray}
E_{x}[\rho] & = & C_{x}\int\rho^{\frac{4}{3}}({\bf r})d{\bf r}\nonumber \\
C_{x} & = & -\frac{3}{4}\left (\frac{3}{\pi}\right )^{\frac{1}{3}}.
\label{49}
\end{eqnarray}
The correlation energy functional within the ALDA is represented by the 
Gunnarsson-Lundquist (GL) parametrization \cite{gl}. Thus
\begin{equation}
E_{c}[\rho] = \int\epsilon_{c}(\rho)\rho({\bf r})d{\bf r}
\end{equation}
with
\begin{equation}
\epsilon_{c}(\rho )  =  -0.0333\left [(1 + x^{3})\ln(1 + \frac{1}{x}) +
\frac{1}{2}x -x^{2} -\frac{1}{3}\right ] 
\end{equation} 
where 
\begin{equation}
x = \frac{r_{s}}{11.4}
\end{equation}
and $r_{s} = (\frac{3}{4\pi}\frac{1}{\rho})^{\frac{1}{3}}$ measures the radius
in atomic units of a sphere which encloses one electron.

The theory presented here treats the non-interacting KE exactly for single orbital systems 
(hydrogen and helium atoms). We have checked this by calculating the 
linear and nonlinear polarizabilities of H and He atoms.
They match well with the corresponding wavefunctional results. The real test 
of the theory is therefore when it is applied to systems with more than one 
orbitals. We now present the results of these calculations by first 
discussing the frequency dependent polarizability numbers for the inert gas 
atoms of neon and argon. The ground-state electronic densities of these atoms
are obtained by employing the van Leeuwen and Baerends (LB) \cite{lb} 
potential. We use this potential as the orbitals generated by it have the 
correct asymptotic nature so that they lead to accurate
values for the static \cite{baner5} and frequency dependent \cite{baner3}
response properties. 
Here, instead of using the orbitals, we are using the ground-state densities 
generated by this potential.

In Figs. 1 and 2 we show the linear polarizabilities $\frac{\alpha(\omega)}
{\alpha(0)}$ for neon and argon atoms, respectively, as a function of 
$\omega$, obtained by hydrodynamic approach. We compare these results 
(represented by open squares) with those obtained \cite{baner3} within the 
Kohn-Sham formalism shown in the figures by filled squares.
As is evident, the frequency dependence matches quite well with the KS 
approach. This along with the zero frequency results demonstrate that the 
hydrodynamic theory is capable of giving reasonable estimate of 
dynamic polarizabilities in the optical range.  Notice though that
in the present approach the increase of 
$\alpha(\omega)$ with respect to $\omega$ is slightly less.  

To further quantify our results, we have fitted the frequency dependent 
polarizabilities with the formulae \cite{leonard}
\begin{equation}
\alpha(\omega) = \alpha(0)\left (1 + C_{2}\omega^{2}\right )
\label{50}
\end{equation}
for small frequencies (up to $\omega = 0.05$ a.u.). In Table I we give the 
results for $C_{2}$ obtained from both the hydrodynamic and the orbital 
based calculations \cite{baner3}. As anticipated from the discussion above, 
the values of C$_{2}$ obtained from hydrodynamic formulation are close to but
slightly smaller than their wavefunctional counterparts.

Next, to study the performance of hydrodynamic approach in calculation of 
nonlinear response properties we calculate the coefficient corresponding to 
the DFWM phenomenon.  These results  are presented in Table II for two 
different frequencies, $\lambda = 10550$\AA \ ($\omega\approx 0.0433$ a.u.) 
and $\lambda = 6943$\AA \ ($\omega\approx 0.0657$ a.u.).
Here also we compare the present results with the corresponding numbers 
obtained \cite{baner3} by the TD Kohn-Sham method.
From this Table  we again observe that the results for DFWM
coefficients at both the frequencies are also quite close to, but lower than, 
the corresponding wavefunctional numbers. Notice that the numbers for 
hyperpolarizabilities are also underestimated by the hydrodynamic approach 
and the maximum deviation from the TD Kohn-Sham number is about 10\%.  
However, with the increase in frequency the difference between the 
hydrodynamical and the wavefunctional results will get larger. Nonetheless, 
the results obtained are reasonably accurate for frequencies up to 
$\omega\approx 0.05$ a.u.. This is quite encouraging since the experimental 
measurement of nonlinear coefficients fall within this frequency regime and 
also the numerical effort required for the hydrodynamical calculation is 
substantially less in comparison to the wavefunctional approach.

Motivated by these results, in the next section we apply the hyrodynamic 
approach to calculate frequency dependent response properties and plasmon 
frequencies of metal clusters. As is well known, orbital based calculation 
for such systems are computationally demanding because of the large number 
of orbitals involved as the cluster size grows.
\subsection*{Response properties of metal clusters}
The hydrodynamic approach developed above is particularly useful for systems
where an orbital based theory cannot be applied. Clusters are one such class 
of systems. These are made up of tens to thousands of atoms with properties 
distinct from the bulk properties of the constituent material. 
Further, various properties of clusters evolve in a well defined
manner as their size grows. This was demonstrated by Knight et al. 
\cite{knight1,knight2}in their study of alkali metal clusters. Since then 
metal clusters have been studied \cite{heer,brack1,alonso} quite extensively.
One of the
simplest model that describes average properties of these systems correctly 
is the spherical jellium background model (SJBM) \cite{ekardt,brack1,alonso}.
In small size clusters, Kohn-Sham LDA equations can be easily solved within 
this model. On the other hand, for large clusters one switches over
to the density based theories \cite{brack2,sinder}; the mostly applied one 
has been the extended Thomas-Fermi (ETF) theory \cite{brack2}. In this paper 
also we use the density obtained by the ETF to calculate frequency dependent 
response properties (both linear and nonlinear) of alkali metal clusters 
with the number of atoms up to 5000. In the past, dynamic linear polarizabilities
of these clusters have been studied using the time-dependent Kohn-Sham theory
\cite{ekardt,beck,guet} but because of the difficulties
mentioned above, the size up to which one could go has been limited.

In the ETF method the ground-state density is obtained by minimizing the energy functional
\begin{equation}
E[\rho] = T_{s}^{ETF}[\rho] + E_{H}[\rho] + E_{xc}^{LDA}[\rho] + \int V_{I}({\bf r})\rho({\bf r})d{\bf r}
+ E_{I}
\label{51}
\end{equation}
where $V_{I}$ and $E_{I}$ are the potential and the total electrostatic energy, respectively,
of the ionic background. The functional $T_{s}^{ETF}$ is the non-interacting KE functional
included up to the fourth-order in density gradients. It is given as \cite{dreizler}
\begin{equation}
T_{s}^{ETF}[\rho] =  T_{s}^{(0)}[\rho] + T_{s}^{(2)}[\rho] + T_{s}^{(4)}[\rho]
\label{52}
\end{equation}
where
\begin{eqnarray}
T_{s}^{(0)} & = & (3\pi^{2})^{2/3}\int\rho^{\frac{5}{3}}d{\bf r} \nonumber \\
T_{s}^{(2)} & = & \frac{1}{72}\int\frac{|{\bf\nabla}\rho |^{2}}{\rho}d{\bf r} \nonumber \\
T_{s}^{(4)} & = & \frac{1}{540(3\pi^{2})^{\frac{3}{2}}}\int\rho^{\frac{1}{3}}\left [\left (\frac{\nabla^{2}\rho}{\rho}\right )^{2}
- \frac{9}{8}\left (\frac{\nabla^{2}\rho}{\rho}\right )\left (\frac{{\bf\nabla}\rho}{\rho}\right )
^{2} + \frac{1}{3}\left (\frac{{\bf\nabla}\rho}{\rho}\right )^{4}\right ]d{\bf r},
\label{53}
\end{eqnarray}
E$_{H}[\rho]$ is the Hartree energy and for E$_{xc}^{LDA}$ is the LDA XC 
energy. For this we use
the GL parametrization \cite{gl}. The energy above is minimized by taking 
the variational form \cite{brack2} for the density to be
\begin{equation}
\rho(r) = \rho_{0}\left [ 1 + exp\left (\frac{r - R}{\alpha}\right )\right ]^{-\gamma}
\label{54}
\end{equation}
where $R$, $\alpha$ and $\gamma$ are the variational parameters and $\rho_{0}$ is fixed by 
the normalization condition for each set of these parameters. The density so obtained gives 
results which are quite close \cite{brack2} to the results of more accurate 
KS calculations of several properties.  We use the ground-state densities obtained by this method
as the input for the calculation of response properties. We perform our calculations for sodium
clusters with $r_{s}=4.0$, where $r_{s}$ is Wigner-Seitz radius of metal.
Before presenting our results we point out that that the KE functional used
to obtain the ground-state density and that for calculating the response
properties are different. This is because whereas ETF functional is good
for the total energies, it does not give the changes in the energies 
accurately. As mentioned earlier, for this purpose 
we use the von Weizsacker \cite{von} functional. 

First we present the results for static linear polarizabilities.
Although polarizability of metal clusters has been investigated 
extensively in the past \cite{ekardt,beck,guet}, these studies 
have been restricted to clusters with number of atoms 
up to 200 because of the use of the orbitals in the calculations. We perform 
our study for clusters up to 5000 atoms. The variational forms for the 
induced particle and the current  densities are chosen to be similar to 
those used in the atomic case.
In Fig.3 we show plot of static polarizability $\alpha(0)$ in the units of 
$R_{0}^{3}$ as a function of $R_{0}$ (where $R_{0} = r_{s}N^{\frac{1}{3}}$,
denotes the radius of cluster). It clearly shows that results of our 
calculation match quite well with the results obtained by Kohn-Sham approach
for small (up to 196 atoms) clusters. As the size of cluster grows the 
polarizability approaches the classical limit of $R_{0}^{3}$ (that is
$\frac{\alpha(0)}{R_{0}^{3}}\rightarrow 1$). This is exhibited rather clearly
in Fig.3.

Having obtained the static polarizabilities of alkali metal clusters 
accurately, we next study the dynamic response properties of metal 
clusters focussing our attention particularly on the dipole resonance. 
The classical theory of dynamic polarizability predicts a single dipole 
resonance at the frequency given by \cite{mie} (in a.u.)
\begin{equation}
\omega_{Mie} = \sqrt{\frac{1}{r_{s}^{3}}},
\label{55}
\end{equation}
which is equal to $1/\sqrt{3}$ times the bulk plasma frequency. The TDDFT 
results for the dipole resonance follow \cite{brack1,alonso} the Mie result 
only in a qualitative way. The resonance peak corresponding to the 
photo absorption spectra of these clusters exhaust about 70\%-90\% of the 
dipole sum rule and is red-shifted by about 10\%-20\% from the classical
Mie formula \cite{brack1,alonso}.

In our work we estimate these resonance peaks from the frequency dependent 
polarizabilities by approximately locating the frequency at which 
$\alpha(\omega)$ becomes very large. These results are presented in Table III
along with the results obtained by Brack \cite{brack2} using the RPA sum 
rules. It is clear from Table III that the dipole resonance frequencies of 
metal clusters obtained by hydrodynamical approach to TDDFT are quite 
accurate over the range of clusters
studied. Further, the accuracy is better for larger clusters. We
also find that with the increase in particle size the dipole resonance 
frequency approaches the classical Mie resonance frequency, which in 
the present case of $r_{s} = 4.0$ is 0.125 a.u..

Next we discuss the results obtained for the nonlinear polarizabilities of
metal clusters by the present approach. To the best of our knowledge
hyperpolarizabilities for these systems have not been calculated
before the present work. In these calculations we are 
restricted to clusters with maximum of only 300 atoms due to  
computational difficulties. Since electrons in metallic cluster are highly 
delocalized, we expect that the nonlinear response of these system
should be quite large and increase  
rapidly with the size of the clusters. To ascertain how does the static 
hyperpolarizability $\gamma(0)$ scale with the size of clusters, in 
Fig.4 we plot $\gamma(0)$ versus $\alpha(0)$ on a log-log scale.
The line in this figure represents the best fit to hyperpolarizability 
versus polarizability numbers obtained by us.
It is seen from figure that for the clusters studied by us, the 
hyperpolarizability is linearly proportional to the linear polarizability.
Since $\alpha(0)$ scales linearly with N, 
$\gamma(0)$ also varies in the same manner. This is a surprising result since 
in the atomic case we have seen that $\gamma(0)$ increases much more 
rapidly than $\alpha(0)$ does. This could be because the electrons in metal
clusters are much more mobile and therefore screen the applied potential
very efficiently.

We have also studied the frequency dependent hyperpolarizabilities 
$\gamma(\omega;\omega,-\omega,\omega)$ and found it increasing with $\omega$. 
Variation of $\gamma(\omega;\omega,-\omega,\omega)$ with $\omega$
is shown in Fig. 5. It becomes quite large (by an order of magnitude in
comparison to the static result) at approximately half the dipolar resonance 
frequency obtained from $\alpha(\omega)$ (Table III). This demonstrates the 
inherent consistency of the theory.

Our study above has been done for clusters with number of atoms up to 300. 
However, the trends obtained should continue as the size grows. Slow increase
of $\gamma$ with N is consistent with the fact that the classically $\gamma$
for spherical metal particle is zero. One reason why computation
becomes difficult for large clusters, we suspect, is that the variational
forms chosen for the second-order particle and current densities may not
be appropriate for very large clusters. As such investigations for larger
clusters relegated to the future studies.
\section*{Concluding remarks}
In this paper we have developed the time-dependent perturbation theory for
periodic (in time) hamiltonian in terms of the particle and current densities
of electrons. For this we have employed the hydrodynamic equations of TDDFT.
Application of the theory developed requires that the energy functional be
approximated. We have demonstrated that with the von Weizsacker \cite{von}
functional for the KE and the ALDA for the XC energy, the theory leads to 
reasonably accurate results for dynamic response properties, both linear
and nonlinear of atoms. Having established that we have applied the theory
to study response properties of metallic clusters with number of atoms up
to 5000 within the SJBM. Of particular interest is how the 
hyperpolarizability varies with the size of these clusters. Although
it is zero classically, our study shows that it increases linearly with the 
number of atoms in the cluster.
 
{\bf Acknowledgement}: We thank Prof. M. Brack for sending us his program
to calculate ground-state density using the ETF approach.

\newpage

\newpage
\subsection*{Table Captions}
{\bf Table I} $C_{2}$ for inert gas atoms obtained by using hydrodynamical and
wavefunctional approaches.

{\bf Table II} DFWM coefficient $\gamma(\omega;\omega,-\omega,\omega)$ (in
atomic units) for inert gas atoms using hydrodynamic approach.

{\bf Table III} Estimate of dipole resonance frequencies (in atomic units)
of some metal clusters by using hydrodynamic approach. 
\newpage
\begin{center}
\section*{Table I}
\end{center}
\tabcolsep=0.25in
\begin{center}
\begin{tabular}{|c|c|c|}
\hline
Atoms & $C_{2}$ & $C_{2}$ \\
       & (hydrodynamical)& (wavefunctional)$^{(a)}$\\
\hline
He & 1.12 & 1.12 \\ 
Ne & 0.82 & 1.04  \\ 
Ar & 2.16 & 2.65 \\ 
Kr & 2.79 & -  \\ 
\hline
\end{tabular}
\end{center}
\begin{center}
(a) Ref. \cite{baner3}
\end{center}
\begin{center}
\section*{Table II}
\end{center} 
\tabcolsep=0.1in
\begin{center}
\begin{tabular}{|c|c|c|c|c|}\hline
Atoms & \multicolumn{2}{c|}{$\lambda = 6943\AA$} &
\multicolumn{2}{c|}{$\lambda = 10550\AA$}  \\ \cline{2-5}
& hydrodynamic & wavefunctional$^{(a)}$  & hydrodynamic & wavefunctional$^{(a)}$  \\
\hline

He & 44.47 & 44.57 &  43.50 & 43.58  \\
Ne & 83.38& 94.06 & 82.58 & 91.65  \\
Ar & 1095.2 & 1226.1 & 1039.3 & 1154.8 \\
Kr & 2392.7 & - & 2229.5 & -   \\
Xe & 5821.1 & - & 5295.3 & -  \\
\hline
\end{tabular}
\end{center}
\begin{center}
(a) Ref. \cite{baner3}
\end{center}
\begin{center}
\section*{Table III}
\end{center}
\tabcolsep=0.25in
\begin{center}
\begin{tabular}{|c|c|c|}
\hline
N & Present & RPA$^{(a)}$ \\
\hline
8 & 0.100 & 0.113 \\ 
100& 0.105 & 0.1198  \\ 
500 & 0.1165 & 0.1219 \\ 
1000 & 0.121 & 0.1226  \\
5000 & 0.1223 & 0.1236 \\
\hline
\end{tabular}
\end{center}
\begin{center}
(a) Ref. \cite{brack1}
\end{center}
\newpage
\subsection*{Figure Captions}
{\bf Fig.1} Plot of $\alpha(\omega)/\alpha(0)$ as a function of frequency
$\omega$ for neon. The open and closed squares represent hydrodynamical
and wavefunctional \cite{baner3} results respectively. 

{\bf Fig.2} Plot of $\alpha(\omega)/\alpha(0)$ as a function of frequency
$\omega$ for argon. The open and closed squares represent hydrodynamical
and wavefunctional \cite{baner3} results respectively.

{\bf Fig.3} Static polarizability $\alpha(0)$ in the units of $R_{0}^{3}$
of alkali metal clusters as a function of $R_{0}$. The filled 
squares represents the results of Kohn-Sham calculations \cite{guet}.

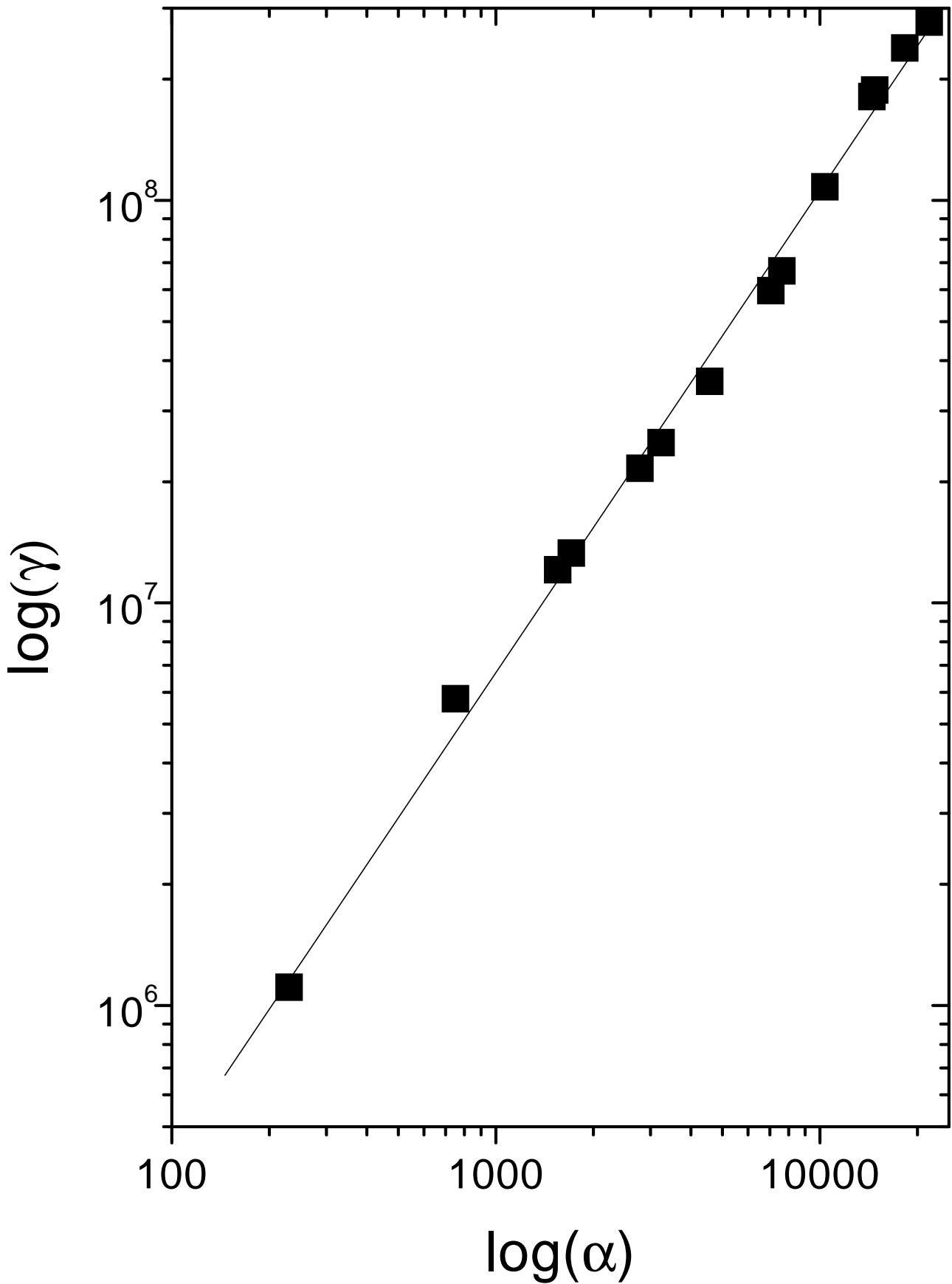
{\bf Fig.4} Plot of $\gamma(0)$ versus $\alpha(0)$ of alkali metal clusters.

{\bf Fig.5} Plot of $\gamma(\omega;\omega,-\omega,\omega)$ in the units of
$R_{0}^{3}$ as a function of frequency $\omega$ for a metal cluster with
100 atoms.
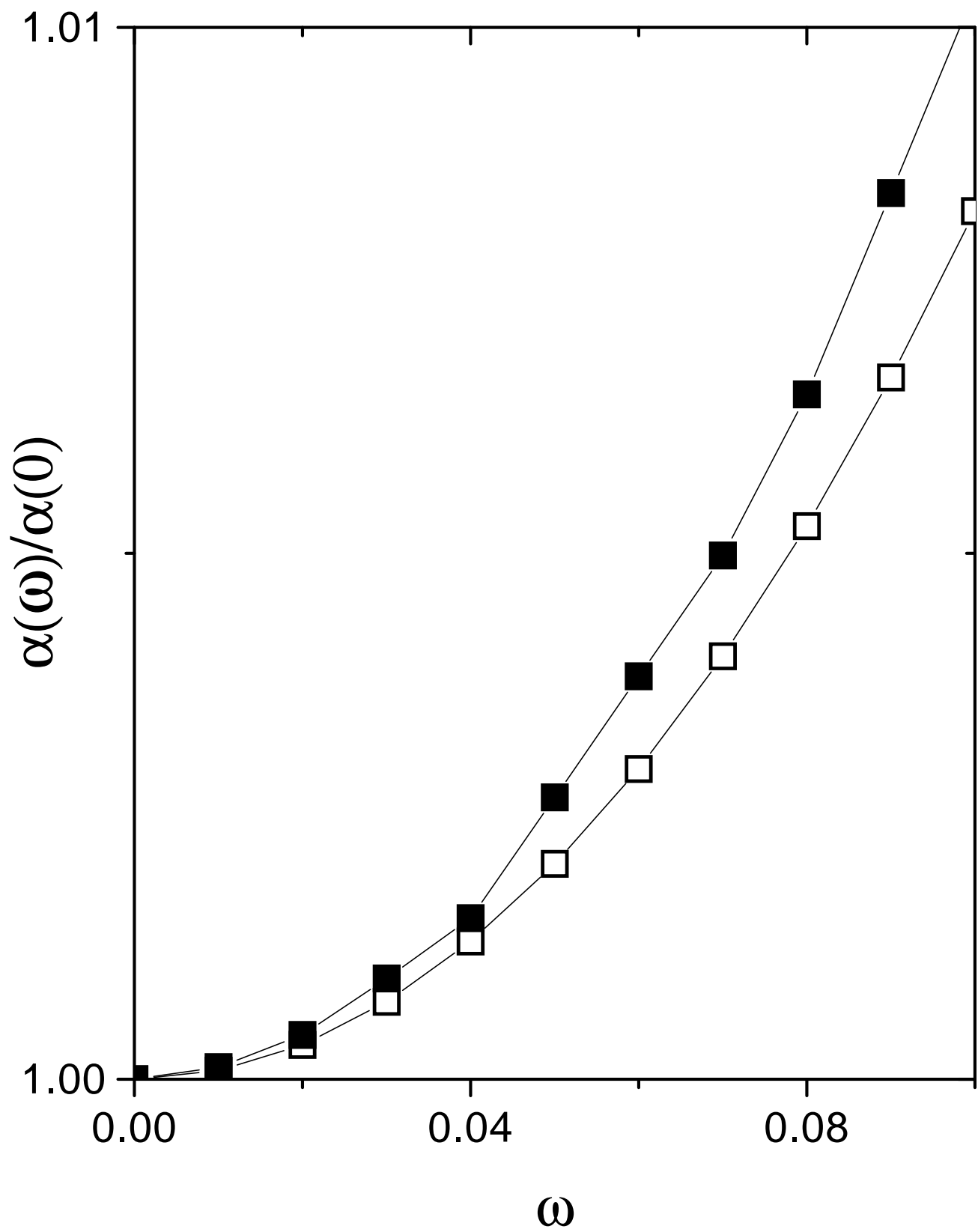
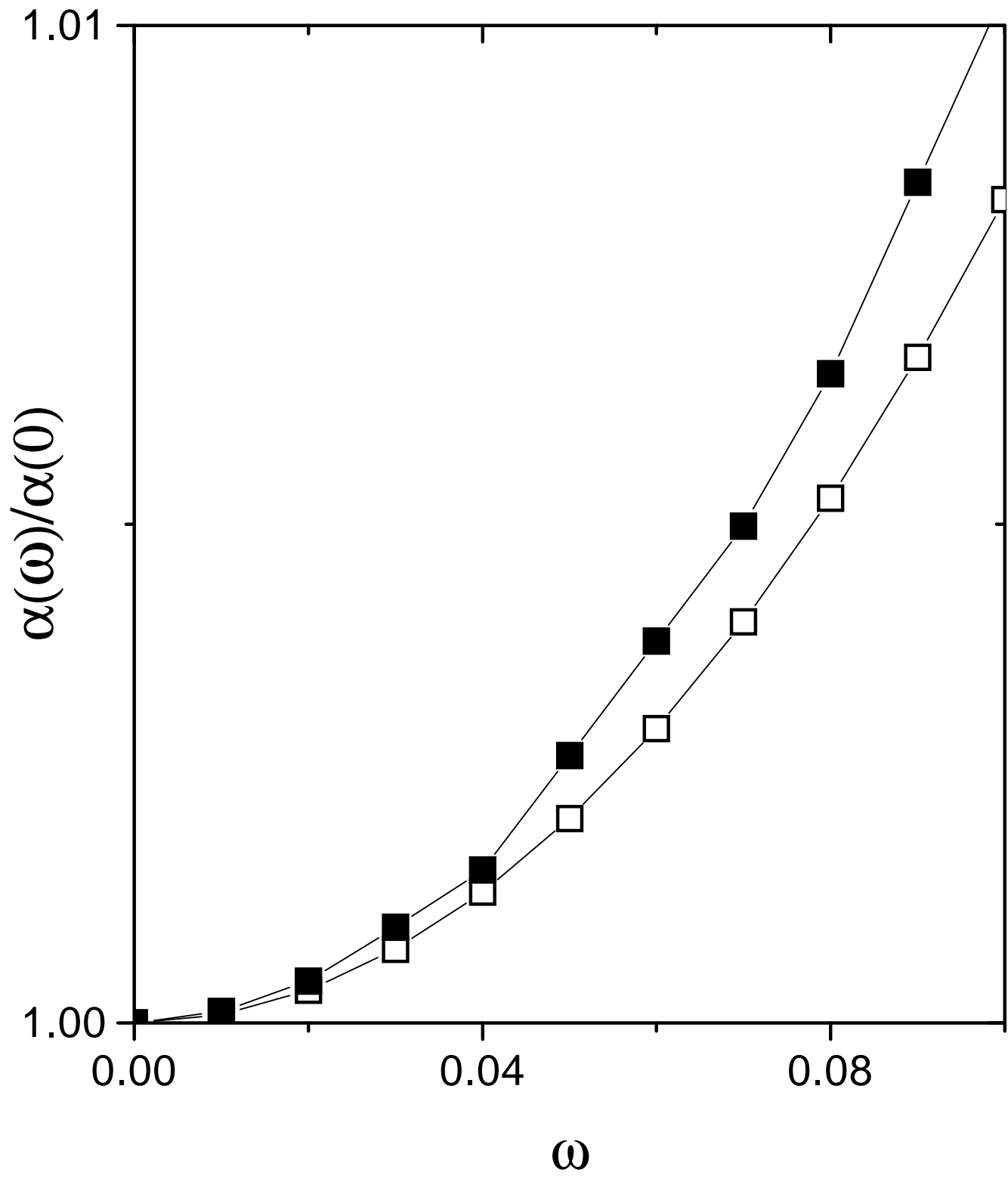
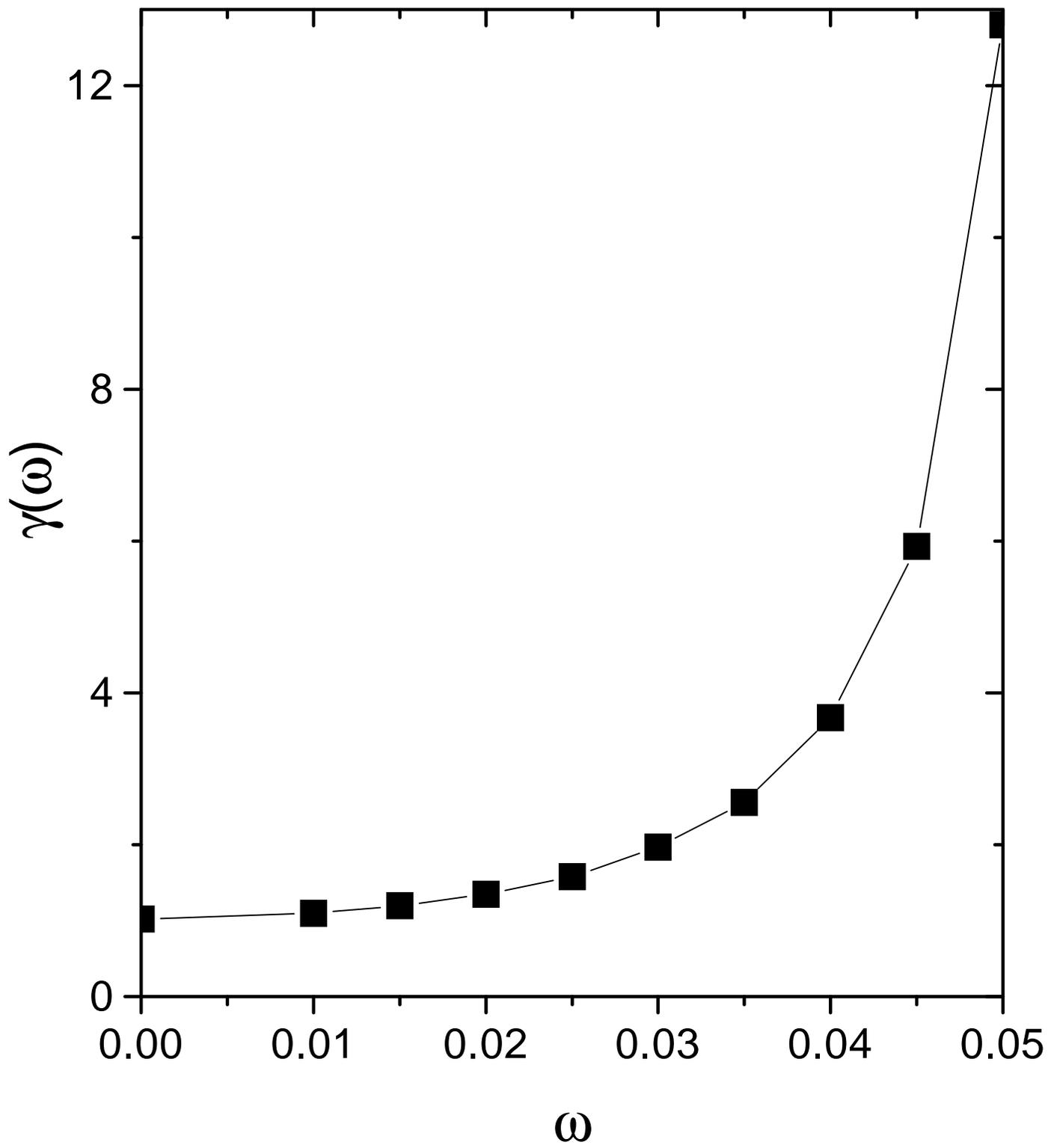
\end{document}